\documentclass[3p,times]{elsarticle}

\usepackage{ecrc}

\volume{preprint}

\firstpage{1}

\journalname{Nuclear Physics A}

\runauth{S. Petschauer}

\jid{NUPHA}

\jnltitlelogo{Nuclear Physics A}

\CopyrightLine{2012}{Published by Elsevier Ltd.}

\usepackage{amssymb, amsmath, mathrsfs, bbm}

\newlength{\feynwidth} 
\DeclareMathOperator{\diag}{diag}
\DeclareMathOperator{\tr}{tr}

\begin{document}

\begin{frontmatter}

\dochead{}

\title{SU(3) chiral dynamics and baryon-baryon interactions}
\author{S. Petschauer}
\address{Physik Department, Technische Universit\"at M\"unchen, D-85747 Garching, Germany}
\ead{stefan.petschauer@ph.tum.de}

\begin{abstract}
We calculate hyperon-nucleon and hyperon-hyperon interactions at next-to-leading order in SU(3) baryon chiral perturbation theory extending earlier work by the Bonn-Juelich group.
The constructed potentials in momentum space include all one- and two-meson exchange terms generated by the SU(3) chiral Lagrangian.
Effects from intermediate decuplet baryons are considered as well.
These chiral baryon-baryon potentials, together with appropriate contact terms, provide a new basis for systematic studies of hyperon-nucleon scattering and light hypernuclei.
\end{abstract}


\end{frontmatter}


\section{Introduction}

Chiral perturbation theory has become a powerful tool for the  systematic calculation of hadronic processes.
At present a very accurate description of low-energy nucleon-nucleon scattering has been achieved in SU(2) chiral effective field theory \cite{Bernard1995,Epelbaum2004,Machleidt2011}.
The extension to the three-flavor case, relevant for describing the baryon-baryon interactions in all (strangeness and isospin) channels has so far not been treated in that detail, also due to the present shortage of experimental scattering data.
A leading order (i.e.\ tree level) calculation of hyperon-nucleon scattering has been performed in chiral perturbation theory in Ref.~\cite{Polinder2006}.
In this work we consider the two-meson-exchange contributions to the baryon-baryon potentials at next-to-leading order.
These chiral hyperon-nucleon potentials together with chiral contact terms are basic input for calculations of baryon-baryon scattering, hypernuclei and strange baryonic matter.

\section{Theoretical framework}

In the following we show the relevant parts of the effective chiral Lagrangian.
The leading order purely mesonic Lagrangian reads
\begin{equation}
  \mathscr{L}_\phi^{(2)}=\frac{f_0^2}4 \tr\left(\partial_\mu U \partial^\mu U^\dagger\right) + \frac12 B_0 f_0^2 \tr\left(M U^\dagger+U M\right) \,,
\end{equation}
with the quark mass matrix \(M=\diag\left(m_u,m_d,m_s\right)\) and \(f_0\) the pseudoscalar meson decay constant in the chiral limit.
We use the Lagrangian in the isospin limit which leads to explicit chiral and flavor symmetry breaking by different meson masses in the resulting propagators of the mesons.
The leading order meson-baryon interaction Lagrangian is given by
\begin{equation}
 \mathscr{L}_\mathrm{MB}^{(1)}=\tr\left(\bar B \left(\mathrm i \gamma^\mu D_\mu -M_0\right) B\right) - \frac D 2 \tr\left(\bar B \gamma^\mu \gamma_5 \lbrace u_\mu,B\rbrace\right) - \frac F 2 \tr\left(\bar B \gamma^\mu \gamma_5 [u_\mu,B]\right)\,,
\end{equation}
with
\( D_\mu B = \partial_\mu B + [\Gamma_\mu,B]\),
\( \Gamma_\mu = \frac 1 2 ( u^\dagger \partial_\mu u + u \partial_\mu u^\dagger )\) and
\(u_\mu = \mathrm i ( u^\dagger \partial_\mu u - u \partial_\mu u^\dagger )\).
The constant \(M_0\) is the baryon mass in  the chiral limit.
For the pseudoscalar mesons and octet baryons
\begin{equation}
 \phi=
 \begin{pmatrix}
  \pi^0 + \frac{\eta}{\sqrt 3} & \sqrt 2 \pi^+ & \sqrt 2 K^+ \\
  \sqrt 2 \pi^- & -\pi^0 + \frac{\eta}{\sqrt 3} & \sqrt 2 K^0 \\
  \sqrt 2 K^- & \sqrt 2 \bar K^0 & -\frac{2\eta}{\sqrt 3}
 \end{pmatrix}\,,\qquad
  B=
 \begin{pmatrix}
  \frac{\Sigma^0}{\sqrt 2} + \frac{\Lambda}{\sqrt 6} & \Sigma^+ & p \\
  \Sigma^- & -\frac{\Sigma^0}{\sqrt 2} + \frac{\Lambda}{\sqrt 6} & n \\
  \Xi^- & \Xi^0 & -\frac{2\Lambda}{\sqrt 6}
 \end{pmatrix}\,,
\end{equation}
we use the usual non-linear realization of chiral symmetry with \(U(x) = u^2(x)=\exp\left(\mathrm i\phi(x)/f_0\right)\).
These fields transform under the chiral symmetry group \(\mathrm{SU}(3)_\mathrm L \times \mathrm{SU}(3)_\mathrm R\) as
\( U \rightarrow RUL^\dagger\) and
\(B \rightarrow K B K^\dagger\)
with \(L\in\mathrm{SU}(3)_\mathrm L\,, R\in\mathrm{SU}(3)_\mathrm R\) and the SU(3) valued compensator field \(K=K(L,R,U)\).
The leading order interaction Lagrangian including decuplet baryons is given in the non-relativistic limit by
\begin{equation}
 \mathscr{L}^{(1)}_\mathrm{MBD} = \frac C{f_0} \sum_{a,b,c,d,e=1}^3 \epsilon_{abc} \left( \bar T_{ade} \vec S^{\,\dagger} \cdot \left(\vec\nabla \phi_{db}\right)B_{ec} + \bar B_{ce} \vec S \cdot \left(\vec\nabla\phi_{bd}\right)T_{ade} \right)\,.
\end{equation}
The spin transition matrix \(\vec S\) connects the two-component spinors of octet baryons with the four-component spinors of decuplet baryons. The operator \(\vec S\) fulfills the relation \( S_i {S_j}^\dagger = \frac13 ( 2\delta_{ij}-\mathrm i\epsilon_{ijk} \sigma_k )\).
The decuplet baryons are represented by the totally symmetric three-index tensor \(T\),
\begin{equation}
  \begin{aligned}
  T^{111}&=\Delta^{++}\,,
 \quad& T^{112}&=\tfrac{1}{\sqrt{3}}\Delta^{+}\,,
 \quad& T^{122}&=\tfrac{1}{\sqrt{3}}\Delta^{0}\,,
 \quad& T^{222}&=\Delta^{-}\,,
 \quad& T^{113}&=\tfrac{1}{\sqrt{3}}\Sigma^{*+}\,,\\
 T^{123}&=\tfrac{1}{\sqrt{6}}\Sigma^{*0}\,, & T^{223}&=\tfrac{1}{\sqrt{3}}\Sigma^{*-}\,,
 \quad& T^{133}&=\tfrac{1}{\sqrt{3}}\Xi^{*0}\,, & T^{233}&=\tfrac{1}{\sqrt{3}}\Xi^{*-}\,,
 \quad& T^{333}&=\Omega^-\,,
  \end{aligned}
\end{equation}
which transforms under chiral symmetry as \(T_{abc} \rightarrow K_{ad}K_{be}K_{cf}T_{def}\).

\begin{figure}[t]
 \centering
 \includegraphics[width=\feynwidth]{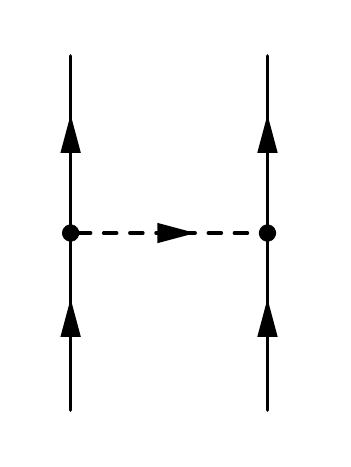}
 \includegraphics[width=\feynwidth]{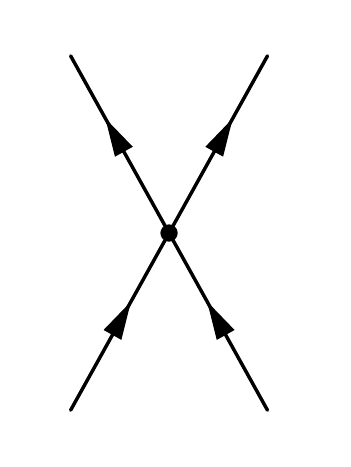}
 \includegraphics[width=\feynwidth]{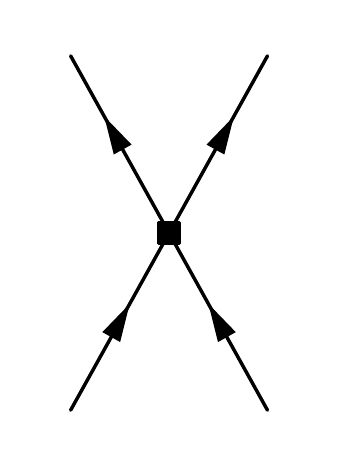}
 \includegraphics[width=\feynwidth]{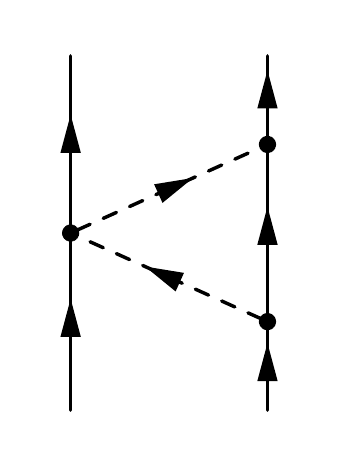}
 \includegraphics[width=\feynwidth]{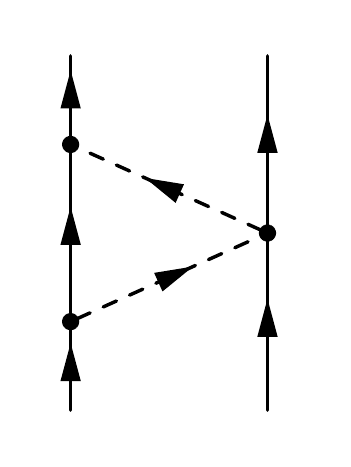}
 \includegraphics[width=\feynwidth]{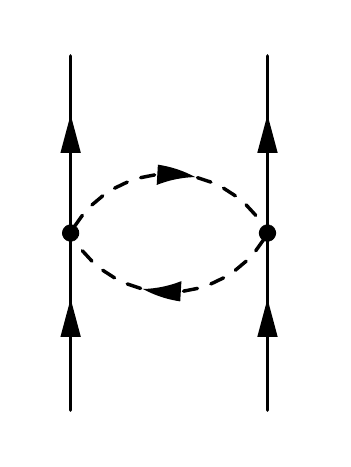}
 \includegraphics[width=\feynwidth]{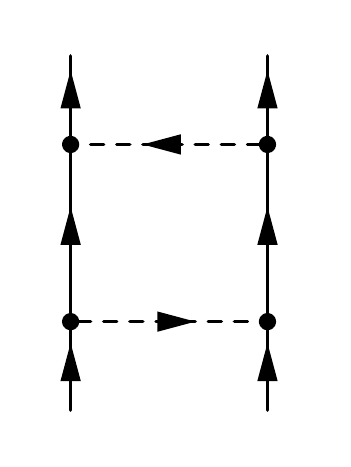}
 \includegraphics[width=\feynwidth]{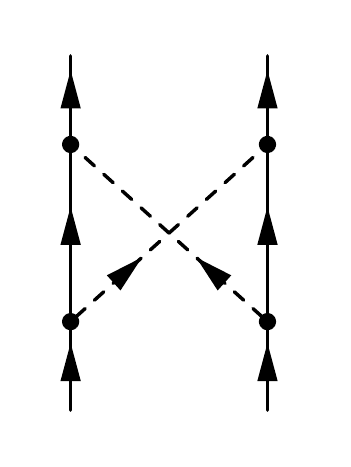} \\[-.6\baselineskip]
 \includegraphics[width=\feynwidth]{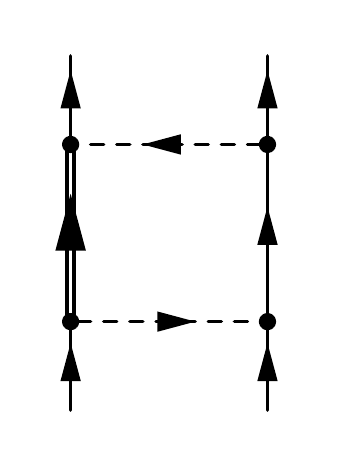}
 \includegraphics[width=\feynwidth]{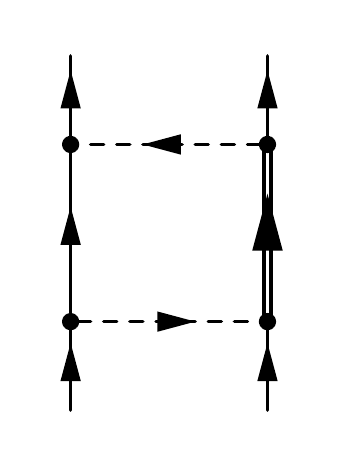}
 \includegraphics[width=\feynwidth]{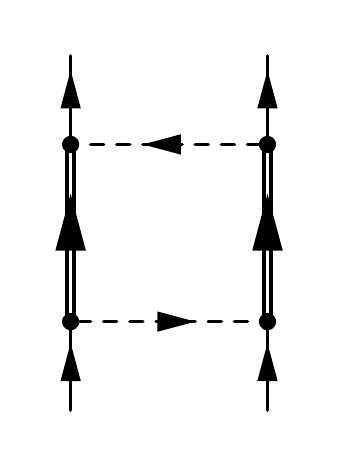}
 \includegraphics[width=\feynwidth]{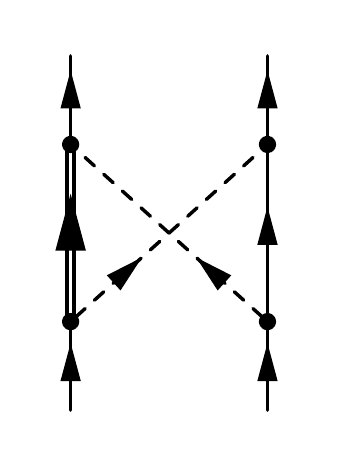}
 \includegraphics[width=\feynwidth]{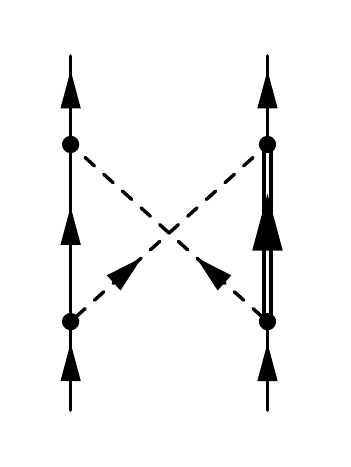}
 \includegraphics[width=\feynwidth]{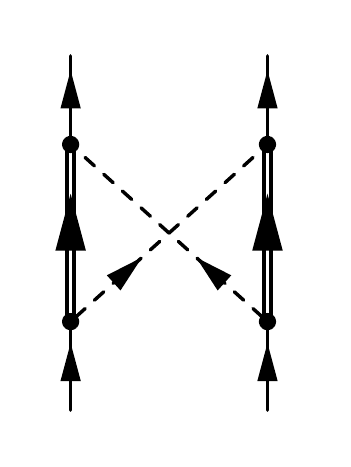}
 \includegraphics[width=\feynwidth]{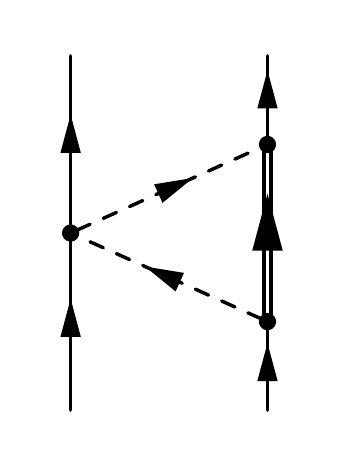}
 \includegraphics[width=\feynwidth]{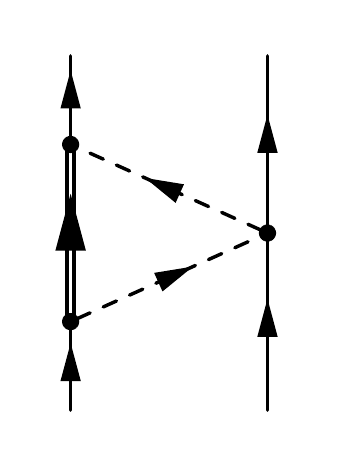}
 \caption{Possible Feynman diagrams up to NLO. Solid, double and dashed lines are octet, decuplet baryons and mesons, respectively. Dots are vertices proportional to \(q^0\). A square stands for a vertex proportional to \(q^2\).} \label{fig:pwrcounting}
\end{figure}

We follow the power counting scheme of Weinberg, where one defines the potential as the two-particle irreducible part of the T-matrix, and obtains the full T-matrix by iterating this potential with a Lippmann-Schwinger equation to all orders.
The potentials of baryon-baryon interactions are ordered in powers of small momenta according to \cite{Polinder2006}
\begin{equation}
 V_{\mathrm{eff}}=V_{\mathrm{eff}}\left(q,g,\mu\right)=\sum_\nu q^\nu \mathcal{V}_\nu\left(q/\mu,g\right)\,,
\end{equation}
with the chiral dimension \(\nu=2-\tfrac12 B+2L+\sum_i v_i \Delta_i\) and \(\Delta_i=d_i+\tfrac12 b_i-2\).
The number of external baryons is denoted by \(B\) and \(L\) is the number of Goldstone boson loops,
\(v_i\) is the number of vertices with dimension \(\Delta_i\).
For a vertex with dimension \(\Delta_i\), the number of derivatives or Goldstone boson masses is denoted by \(d_i\), and \(b_i\) is the number of internal baryon lines.
The soft scale \(q\) is either a baryon three-momentum, a Goldstone boson four-momentum or a Goldstone boson mass.
Following this scheme the leading order (\(\nu=0\)) potential is given by one-meson-exchange diagrams and non-derivative four-baryon contact terms.
At next-to-leading order (\(\nu=2\)) higher order contact terms and two-meson-exchange diagrams with intermediate octet or decuplet baryon contribute.
We use the heavy-baryon formalism where a \(1/M_0\) expansion is performed before doing the loop integrations.
In Fig.~\ref{fig:pwrcounting} we show the possible Feynman diagrams up to next-to-leading order.
In the particle basis different assignments of particles to the lines change only an SU(3) factor and the involved masses, but otherwise the result of the Feynman diagram stays the same.
For diagrams with exchanged (crossed) final state baryons (e.g. YN interaction diagrams with strangeness exchange) one has additionally to apply the spin exchange operator, and has to exchange the momenta of the final state baryons.

\section{Results}

As a representative example we show the potential for the planar box diagram.
It has an irreducible part and a reducible part coming from iterating the one-meson exchange to second order.
One obtains a central potential (\(\mathbbm1\)), a spin-spin potential (\(\vec\sigma_1\cdot\vec\sigma_2\)) and a ``tensor'' potential (\(\vec\sigma_1\cdot\vec q\;\vec\sigma_2\cdot\vec q\,\)).
In the center-of-mass frame, with the momentum transfer \(q=\left|\vec p^{\,\prime} - \vec p\,\right|\) and the masses of the two exchanged mesons, \(m_1\) and \(m_2\), we get the irreducible potentials in closed analytical form,
\begin{align}
 V^\text{box}_\mathrm{irr,\,C}(q) =&-\frac{N}{3072 \pi ^2 f_0^4}\Bigg[\frac{5}{3}q^2+16 \left(m_1^2+m_2^2\right)
 +\frac{\left(m_1^2-m_2^2\right)}{q^4} \left(12 q^4+\left(m_1^2-m_2^2\right)^2-9 q^2 \left(m_1^2+m_2^2\right)\right) \ln \frac{m_1}{m_2}\notag\\
 &+\frac{\left(m_1^2-m_2^2\right)^2}{q^2}+\left(46 q^2+90 \left(m_1^2+m_2^2\right)\right) \ln \frac{\sqrt{m_1 m_2}}{\lambda }+\frac{2}{w^2\left(q\right)} \bigg(23 q^4-\frac{\left(m_1^2-m_2^2\right)^4}{q^4}+56 \left(m_1^2+m_2^2\right) q^2\notag\\
 &+8\frac{ m_1^2+m_2^2}{q^2}\left(m_1^2-m_2^2\right)^2 + 2 \left(21 m_1^4+22 m_1^2 m_2^2+21 m_2^4\right)\bigg) L\left(q\right)\Bigg]\,,
\end{align}
\begin{equation}
 V^\text{box}_\mathrm{irr,\,T}\left(q\right) = \frac{N}{128 \pi ^2 f_0^4} \Bigg(L\left(q\right)-\frac{1}{2}-\frac{m_1^2-m_2^2}{2 q^2}\ln \frac{m_1}{m_2}
 +\ln\frac{\sqrt{m_1 m_2}}{\lambda }\Bigg) = -\frac1{q^2}V^\text{box}_\mathrm{irr,\,S}\left(q\right) \,.
\end{equation}
The prefactor \(N\) comprises the coupling constants from the four vertices and it is different for each combination of baryons and mesons in the particle basis.
Ultraviolet divergences are treated by dimensional regularization which introduces the scale \(\lambda\).
Note that divergent terms beyond \(\ln\lambda\) have been dropped.
We have defined the functions
\begin{equation}
 w\left(q\right) = \frac1q\sqrt{\left(q^2+\left(m_1+m_2\right)^2\right)\left(q^2+\left(m_1-m_2\right)^2\right)}\,,\qquad
 L\left(q\right) = \frac{w\left(q\right)}{2q} \ln \frac{\left( qw\left(q\right) + q^2\right)^2 - \left(m_1^2-m_2^2\right)^2}{4m_1m_2q^2}\,.
\end{equation}
For the nucleon-nucleon interaction with equal meson masses the expressions reduce to the results in Ref.~\cite{Kaiser1997}.

Similarly, we show the potential for the triangle diagram with an intermediate decuplet baryon.
One obtains the following contribution to the central potential
\begin{align}
 V^\text{dec}_\mathrm{C}(q) =& \frac{C^2N}{576 \pi ^2 f_0^4} \Bigg[\left(6 \Sigma\left(q\right) -w^2\left(q\right)\right) L\left(q\right)+12 \Delta ^2 \Sigma\left(q\right)  D\left(q\right)
 +18 \Delta ^2-\left(m_1^2+m_2^2\right)\notag\\
 &-\frac{13 }{6}q^2+\frac{\left(m_1^2-m_2^2\right)^2}{2 q^2} +\left(9 \left(m_1^2+m_2^2\right)+5 q^2-36 \Delta ^2\right) \ln \frac{\sqrt{m_1 m_2}}{\lambda }\notag\\
 &+\frac{\left(m_1^2-m_2^2\right)}{2 q^4} \left(\left(m_1^2-m_2^2\right)^2-3 q^2 \left(m_1^2+m_2^2\right)+12 q^2 \Delta ^2\right) \ln \frac{m_1}{m_2}\notag\\
 &+12 \Delta  \left(\sqrt{m_1^2-\Delta ^2} \arccos\frac{\Delta }{m_1}+\sqrt{m_2^2-\Delta ^2} \arccos\frac{\Delta }{m_2}\right)\Bigg]\,.
\end{align}
Here we have introduced the functions \(\Sigma(q)\) and \(D(q)\) by
\begin{equation}
 \Sigma(q) = m_1^2+m_2^2+q^2-2\Delta^2\,,\quad
 D(q) = {\frac1\Delta \int\limits^\infty_{m_1+m_2} \!\! \mathrm d\mu^\prime\, \frac1{\mu^{\prime2}+q^2}\arctan\frac{\sqrt{\left(\mu^{\prime2}-\left(m_1+m_2\right)^2\right)\left(\mu^{\prime2}-\left(m_1-m_2\right)^2\right)}}{2\Delta \mu^\prime}}\,.
\end{equation}
The average mass difference between decuplet and octet baryons is denoted by \(\Delta\).
For the nucleon-nucleon interaction this potential reduces to the result given in Ref.~\cite{Kaiser1998}.

\begin{figure}[t]
\centering
\includegraphics[width=\feynwidth]{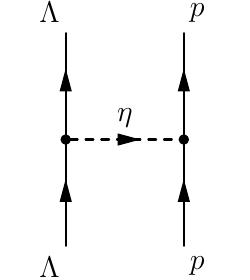}
\includegraphics[width=\feynwidth]{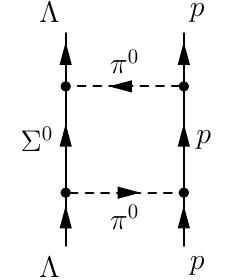}
\includegraphics[width=\feynwidth]{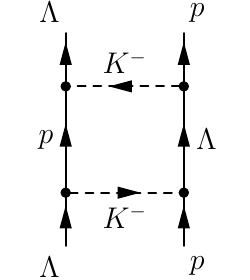}
\includegraphics[width=\feynwidth]{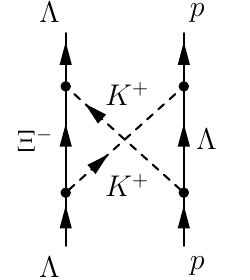}
\includegraphics[width=\feynwidth]{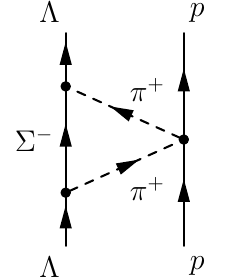}
\includegraphics[width=\feynwidth]{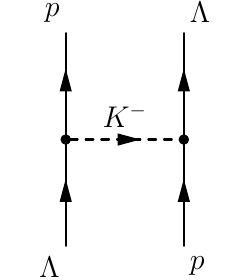}
\includegraphics[width=\feynwidth]{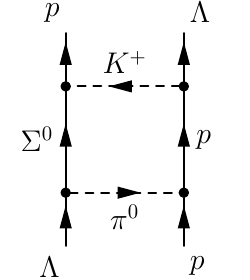}
\includegraphics[width=\feynwidth]{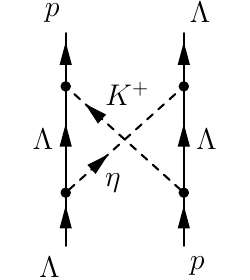} \\[.5\baselineskip]
\includegraphics[width=\feynwidth]{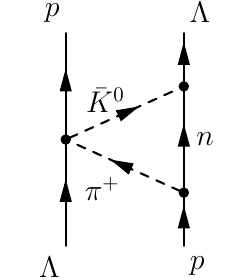}
\includegraphics[width=\feynwidth]{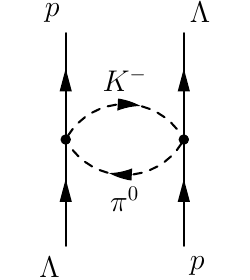}
\includegraphics[width=\feynwidth]{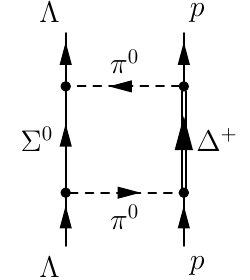}
\includegraphics[width=\feynwidth]{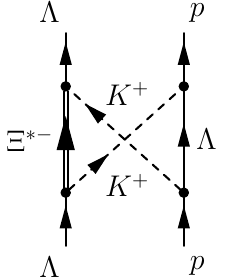}
\includegraphics[width=\feynwidth]{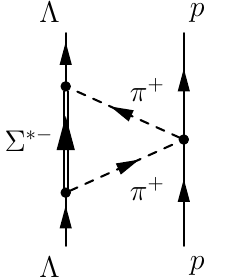}
\includegraphics[width=\feynwidth]{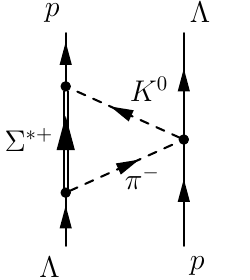}
\includegraphics[width=\feynwidth]{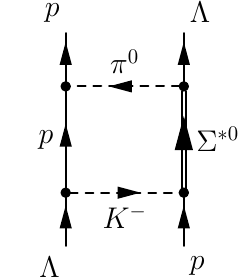}
\includegraphics[width=\feynwidth]{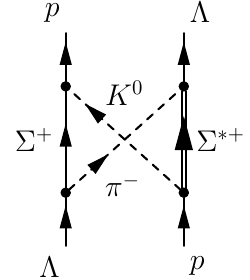}
\caption{Selection of Feynman diagrams contributing to the transition \(\Lambda p \rightarrow \Lambda p\).} \label{fig:lplp-ex}
\end{figure}

For the hyperon-nucleon interaction with total isospin \(1/2\) and strangeness \(-1\) one has to deal with the coupled channels \(\Lambda N\) and \(\Sigma N\).
For example, in the diagonal part \(\Lambda N \rightarrow \Lambda N\) one has to evaluate 93 contributing Feynman diagrams.
Fig.~\ref{fig:lplp-ex} shows a representative set of these one- and two-meson-exchange diagrams.

\section{Summary and Conclusions}

We have constructed the baryon-baryon potentials in momentum space up to next-to-leading order in SU(3) baryon chiral perturbation theory including one-meson exchange and two-meson exchange diagrams generated by the SU(3) chiral Lagrangian.
Effects from intermediate decuplet baryons are considered as well.
These potentials are to be supplemented by appropriate counter terms which encode the unresolved short-distance dynamics.
Inserting the resulting full potentials into a regularized Lippmann-Schwinger equation, leads with SU(3) symmetric low-energy constants already to a good description of the available hyperon-nucleon scattering cross sections. Detailed results herefore have been presented by J. Haidenbauer \cite{HaidenbauerProc}.
The application of these potentials to light hypernuclei has been reviewed by A. Nogga \cite{NoggaProc}.

\section{Acknowledgements}

I thank my collaborators Johann Haidenbauer, Norbert Kaiser, Ulf-G. Mei\ss{}ner, Andreas Nogga and Wolfram Weise.
This work is supported by the TUM Graduate School (TUM-GS) and by DFG and NSFC (CRC 110).


\end{document}